\documentclass[12pt,draftcls,onecolumn]{IEEEtran}

\def \be {\begin{equation}}
\def \ee {\end{equation}}
\def \mc {\mathcal}
\def \nn {\nonumber}

\usepackage{ifpdf}

\usepackage[dvips]{graphicx}

\usepackage[cmex10]{amsmath}
\usepackage{amssymb}
\begin{document}
\title{Second-Order Consensus of Networked Mechanical Systems With Communication Delays}

\author{Hanlei~Wang and Long Cheng%\emph{IEEE Member}
        % <-this % stops a space
%\thanks{This work was supported by the National Natural Science Foundation of China under Grants 61374060 and 61370032.}%, and the National Key Basic Research Program (973) of China under Grant 2013CB733100.}
\thanks{H. Wang is with Science and Technology on Space Intelligent Control Laboratory,
Beijing Institute of Control Engineering,
Beijing 100190, China (e-mail: hlwang.bice@gmail.com).}
\thanks{L. Cheng is with State Key Laboratory of Management and Control for
Complex Systems, Institute of Automation, Chinese Academy of Sciences,
Beijing 100190, China (e-mail: chenglong@compsys.ia.ac.cn).}% <-this % stops a space
}
\maketitle

\begin{abstract}
%\boldmath
In this paper, we consider the second-order consensus problem for networked mechanical systems subjected to nonuniform communication delays, and the mechanical systems are assumed to interact on a general directed topology. We propose an adaptive controller plus a distributed velocity observer to realize the objective of second-order consensus. It is shown that both the positions and velocities of the mechanical agents synchronize, and furthermore, the velocities of the mechanical agents converge to the scaled weighted average value of their initial ones. We further demonstrate that the proposed second-order consensus scheme can be used to solve the leader-follower synchronization problem with a constant-velocity leader and under constant communication delays. Simulation results are provided to illustrate the performance of the proposed adaptive controllers.
\end{abstract}

\begin{keywords}
Second-order consensus, communication delay, networked mechanical systems, uncertainties, adaptive control.
\end{keywords}

% \fi
\IEEEpeerreviewmaketitle

\section{Introduction}

Synchronization of networked mechanical systems (e.g., robot manipulators, spacecraft, and mobile robots) has received intensive attention in recent years due to its ubiquitous applications in our physical world (see, e.g., \cite{Chopra2006,Spong2007,Chung2009_TRO,Nuno2011_TAC,Min2012_SCL,Ren2009_IJC,Mei2012_Aut,Wang2013b_Aut,Wang2013_TAC,Wang2013b_TAC}).
The major challenge in extending the now well studied results for linear agents (e.g., \cite{Olfati-Saber2004_TAC,Ren2005_TAC,Lee2007_TAC,Yu2010_Aut,Munz2010_Aut}) to mechanical agents lies in the nonlinearity and additionally the possible parametric uncertainties (see, e.g., \cite{Nuno2011_TAC,Wang2013_TAC,Wang2013b_TAC}). This challenge turns out to be more prominent when
there exist communication delays among the agents, as is demonstrated in \cite{Nuno2011_TAC,Min2012_SCL,Wang2013b_TAC}.

The consensus schemes for multiple mechanical systems can roughly be classified into two categories based on different control objectives. In the first category, e.g., \cite{Spong2007,Ren2009_IJC,Nuno2011_TAC,Min2012_SCL,Wang2013_TAC,Wang2013b_TAC,Abdessameud2013_TAC}, the mechanical agents are ensured to reach position consensus while the velocities of the agents converge to zero, and this kind of synchronizing behavior for second-order agents is also called \emph{rendezvous} in the literature (see, e.g., \cite{Lee2006_ACC}). The second category of results (e.g., \cite{Wang2013b_Aut}) ensures position consensus and at the same time non-zero (in most cases) velocity consensus, i.e., the \emph{second-order consensus} is realized. The leader-follower scheme in \cite{Mei2011_TAC,Mei2012_Aut}  may also be put into this category in that relying on a distributed observer, each agent is ensured to track the velocity of the leader. Yet, these second-order consensus schemes (i.e., \cite{Wang2013b_Aut,Mei2011_TAC,Mei2012_Aut}) rely on the assumption that the communication delays are absent. In the presence of communication delays, many rendezvous control algorithms (e.g., \cite{Wang2013b_TAC,Abdessameud2013_TAC,Liu2012_AAS}) are proposed and are shown to be effective, e.g., delay-independent result and stability guaranteed rendezvous are achieved in \cite{Wang2013b_TAC} under constant communication delay and rendezvous is ensured robustly with respect to time-varying communication delays by the small-gain-like approach in \cite{Abdessameud2013_TAC}. The scheme in \cite{Liu2012_AAS}, under a relatively restrictive condition [i.e., the leader position is constant (static leader), and the delay must be uniform and lower than certain upper bound], achieves leader-follower rendezvous (i.e., the leader velocity is zero) with communication delays. However, it remains unclear about \emph{how to achieve second-order consensus of nonlinear mechanical agents under arbitrary constant communication delays}. Note that the delay-robust scheme in \cite{Liu2012_AAS}, in the case of a dynamic leader, can only ensure the boundedness of the leader-follower synchronization errors and in addition, the delay is required to be uniform and lower than certain upper bound. Indeed, there are some solutions to the second-order consensus for linear agents with exactly known models, e.g., consensus without a leader \cite{Yu2010_Aut,Meng2011_SMC} and leader-follower consensus \cite{Peng2009_PhysicaA,Zhu2010_Aut,Yang2010_EJC,Meng2010_EJC}. The results in \cite{Yu2010_Aut,Meng2011_SMC}, yet, demand the delay to be lower than an upper
bound determined by the graph topology, which is difficult to obtain since it relies on certain global quantities of the graph, and moreover, one cannot increase this upper bound via
tuning the controller parameters. Besides, the control scheme in \cite{Meng2011_SMC} drives the velocities of the agents to zero due to the communication delay, which thus only ensures rendezvous rather than second-order consensus.  The results in \cite{Peng2009_PhysicaA,Zhu2010_Aut,Yang2010_EJC,Meng2010_EJC} do not address the more challenging case of no explicit leader [the challenge is due to the weak stability of a leaderless multi-agent system (see, e.g., \cite{Nuno2011_TAC,Wang2013b_TAC})]. In addition, these results either require the availability of the information of the leader to all followers (e.g., \cite{Zhu2010_Aut}) or the delay be lower than certain upper bound (e.g., \cite{Peng2009_PhysicaA,Zhu2010_Aut,Yang2010_EJC,Meng2010_EJC}).

In this paper, we propose a second-order consensus scheme for networked uncertain mechanical systems subjected to nonuniform communication delays, and the communication topology among the agents is assumed to contain a spanning tree. The proposed second-order consensus scheme consists of a delay-robust distributed observer, which provides a velocity-like quantity as a velocity reference for each mechanical agent, and a nonlinear adaptive controller to handle the nonlinearity and the uncertainty of the mechanical agent. By exploiting the iBIBO (integral-bounded-input bounded-output) property  associated with a network transfer function matrix (see, e.g., \cite{Wang2013b_TAC}), we show that the positions and velocities of the mechanical agents synchronize, and in addition, the velocities of the mechanical agents converge to the \emph{scaled weighted average value} of their initial ones, irrespective of the nonuniform constant communication delays (i.e., the communication delays are allowed to be arbitrary finite constants). Our result extends the delay-robust rendezvous algorithms in \cite{Wang2013b_TAC,Nuno2011_TAC,Abdessameud2013_TAC} (which can only ensure the position consensus of the mechanical agents while the velocities of the agents are driven to zero) to realize \emph{non-zero velocity and position consensus}. We, then, demonstrate that the proposed adaptive controller can be used to achieve asymptotic leader-follower synchronization with a constant-velocity leader irrespective of the constant communication delays, which is in contrast to the results in \cite{Liu2012_AAS,Liu2012_CCC} that can only ensure asymptotic leader-follower rendezvous (i.e., the leader velocity is zero) under the case that the communication delays are uniform.

\section{Preliminaries}

\subsection{Graph Theory}

Let us first introduce the digraph theory \cite{Olfati-Saber2004_TAC,Ren2005_TAC,Godsil2001_Book,Ren2008_Book}. Consider $n$ mechanical systems, and the $i$-th mechanical system is denoted by vertex $i$. Let the set $\mathcal{V}=\left\{1,2,\dots,n\right\}$ denote all the agents, and the edge set $\mathcal{E}\subseteq \mathcal{V}\times\mathcal{V}$ denote the information flow among the agents. Define a weighted adjacency matrix $\mathcal{W}=\left[w_{ij}\right]$ according to the rule that $w_{ij}>0$ if $j\in\mathcal{N}_i$, and $w_{ij}=0$ otherwise, where $\mathcal{N}_i=\left\{j|j\in\mathcal{V},\left(i,j\right)\in\mathcal{E}\right\}$ denotes the set of neighboring agents of agent $i$. A directed graph is said to \emph{have a spanning tree} if there is a vertex $k_0\in\mc{V}$ such that any other vertex of the graph has a directed path to $k_0$. The Laplacian matrix $\mathcal{L}_w=\left[\ell_{w,ij}\right]$ is defined as
\be
\label{eq1}
\ell_{w,ij}=
\begin{cases}
\Sigma_{k=1}^n w_{ik} & \text{if $i=j$}\\
-w_{ij} & \text{otherwise}.
\end{cases}
\ee

Several fundamental properties of $\mathcal{L}_w$ are described by the following lemma.

\emph{Lemma 1 (\cite{Ren2005_TAC,Ren2008_Book}):} If the Laplacian matrix $\mathcal{L}_w$ is associated with a digraph containing a spanning tree, then
 \begin{enumerate}
 \item $\mathcal{L}_w$ has a simple zero eigenvalue, and all the other eigenvalues of $\mathcal{L}_w$ have positive real parts;
 \item $\mathcal{L}_w$ has a right eigenvector $1_n=\left[1, 1 ,\dots , 1\right]^T$ and a non-negative left eigenvector $\gamma=\left[\gamma_1, \gamma_2,\dots, \gamma_n\right]^T$ satisfying $\Sigma_{k=1}^n \gamma_k=1$ associated with its zero eigenvalue, i.e., $\mathcal{L}_w 1_n=0$ and $\gamma^T \mathcal{L}_w=0$;
 \item the entry $\gamma_i>0$ if and only if agent $i$ acts as a root of the graph.
\end{enumerate}

\subsection{Equations of Motion of Mechanical Systems}

The equations of motion of the $i$-th mechanical system can be written as \cite{Slotine1991_Book,Spong2006_Book}
\begin{equation}
\label{eq2}
M_i \left( {q_i } \right)\ddot {q}_i + C_i \left( {q_i ,\dot {q}_i }
\right)\dot {q}_i + g_i \left( {q_i } \right) = \tau _i
\end{equation}
where $q_i \in R^m$ is the configuration variable, $M_i \left( {q_i }
\right) \in R^{m\times m}$ is the inertia matrix, $C_i \left( {q_i ,\dot
{q}_i } \right) \in R^{m\times m}$ is the Coriolis and centrifugal matrix,
$g_i \left( {q_i } \right) \in R^m$ is the gravitational torque, and
$\tau _i \in R^m$ is the control torque
exerted on the system.

Three basic properties of the dynamic model (\ref{eq2}) are listed as
follows \cite{Slotine1991_Book,Spong2006_Book}.

\textit{Property 1:} The inertia matrix $M_i (q_i )$ is symmetric and uniformly positive
definite.

\textit{Property 2: }The Coriolis and centrifugal matrix $C_i (q_i ,\dot {q}_i )$ can be
appropriately determined such that $\dot {M}_i(q_i) - 2C_i(q_i,\dot{q}_i) $ be skew-symmetric.

\textit{Property 3: }The dynamics (\ref{eq2}) depends linearly on a
constant parameter vector $a_i $, which yields
\begin{equation}
\label{eq3}
M_i \left( {q_i } \right)\dot {\zeta } + C_i \left( {q_i ,\dot {q}_i }
\right)\zeta + g_i \left( {q_i } \right) = Y_i \left( {q_i ,\dot {q}_i
,\zeta ,\dot {\zeta }} \right)a_i
\end{equation}
where $Y_i \left( {q_i ,\dot {q}_i ,\zeta ,\dot {\zeta }} \right)$ is the
regressor matrix, $\zeta \in R^m$ is a differentiable vector and
$\dot {\zeta }$ is the time derivative of $\zeta $.

\section{Adaptive Consensus Scheme}

In this section, we will design an adaptive controller to realize the second-order consensus, i.e., $q_i(t)-q_j(t)\to 0$ and $\dot{q}_i(t)-\dot{q}_j(t)\to 0$ as $t\to\infty$, $\forall i,j\in\mathcal{V}$.

\subsection{Distributed Velocity Observer}

For the second-order consensus problem without a leader, although there is not any leader in the network, it is still necessary to provide a velocity reference signal to each agent. This can be achieved by designing a \emph{distributed velocity observer} for the $i$-th mechanical agent as
\begin{align}
\label{eq4}
\dot{{v}}_i=&-\Sigma_{j\in \mathcal{N}_i}b_{ij}\left[{v}_i-{v}_j(t-T_{ij})\right]
\end{align}
where ${v}_i$ denotes an observed signal at the side of the $i$-th agent, the initial condition is given as ${v}_i(0)=\dot{q}_i(0)$, $i=1,2,\dots,n$, $T_{ij}$ is the constant communication delay from agent $j$ to agent $i$, and the weighted adjacency matrix $\mathcal{B}=\left[b_{ij}\right]$ is defined similarly to $\mathcal{W}$, i.e., $b_{ij}>0$ if $j\in\mathcal{N}_i$ and $b_{ij}=0$ otherwise. The signal $v_j(t-T_{ij})$ in (\ref{eq4}) denotes the delayed version of $v_j$, and when $0\le t< T_{ij}$, agent $i$ cannot get information from agent $j$ due to the communication delay, in which case, as is typically done, we set $v_j(t-T_{ij})\equiv 0$.
The property of the distributed observer (\ref{eq4}) can be stated as the following lemma.%, where the first result is from \cite{Tian2008_TAC} and the second is from \cite{Wang2013b_TAC}.

\emph{Lemma 2:}
If the graph contains a spanning tree, 1) the observed signals ${v}_i$, $i=1,2,\dots,n$ generated by (\ref{eq4}) are bounded and moreover converge to a common constant value $\bar v$ irrespective of the communication delay, and 2) the consensus value $\bar v$ can be explicitly written as
\be
\label{eq5}
\bar{v}=\frac{1}{1+\Sigma_{k=1}^n\Sigma_{j\in{\cal N}_k}\gamma_k b_{kj}T_{kj}}\Sigma_{k=1}^n\gamma_k\dot{q}_k(0).
\ee

The proof of \emph{Lemma 2} follows immediately from \cite{Tian2008_TAC} and \cite{Wang2013b_TAC}.
 From \emph{Lemma 2}, we see that the observer (\ref{eq4}) can supply each mechanical agent with a velocity reference ${v}_i$ which asymptotically reaches the scaled weighted average value of the initial velocities of the mechanical agents (in the sense of \cite{Wang2013b_TAC}). %Also note that $\bar v$ is actually a constant.

\subsection{Consensus Scheme}

Let us design a delay-dependent reference velocity as
\be
\label{eq6}
\dot{q}_{r,i}=\left(1+{\Sigma_{j\in{\cal N}_i}w_{ij}T_{ij}}\right){v}_i-\Sigma_{j\in{\cal N}_i} w_{ij}\left[q_i-q_j(t-T_{ij})\right]
\ee
and define a sliding vector as
\be
\label{eq7}
s_i=\dot{q}_i-\dot{q}_{r,i}
\ee
where $q_j(t-T_{ij})$ is set to be zero when $0\le t< T_{ij}$, similar to the previous case.

The delay-dependent factor $\Sigma_{j\in\mathcal{N}_i}w_{ij}T_{ij}$ in (\ref{eq6}) is inspired by the leader-follower scheme in \cite{Zhu2010_Aut}, yet, we consider here the more challenging scenario that there is not an explicit leader and only the neighboring information is available, and in addition, as will be shown, the delays are allowed to be any finite constants.

\emph{Remark 1:} In the definition of the reference velocity $\dot q_{r,i}$, we assume that the communication delays are known and constant, yet, in practice, the delays are often time varying and unknown (but bounded). Fortunately, via postponing the use of the received data
 up to the worst-case maximum delays (using the time-stamping technique), the apparent/virtual delays can still be rendered constant and
even exactly known (see, e.g., \cite{Kosuge1996_IROS,Lee2006_TRO}). This may seem somewhat conservative
but shall be acceptable in practice. %Since the proposed
%control allows any finite constant delay (as will be demonstrated below), this technique
%should always be applicable.

Since $\bar v$ is constant, equation (\ref{eq7}) can thus be rewritten as
\begin{align}
\label{eq8}
s_i=&\dot{q}_i-\bar{v}+\Sigma_{j\in\mc{N}_i}w_{ij}\left[q_i-\bar{v}\cdot t-\left(q_j(t-T_{ij})-\bar{v}\cdot(t-T_{ij})\right)\right]\nn\\
&-\left(1+\Sigma_{j\in\mc{N}_i}w_{ij}T_{ij}\right)\left({v}_i-\bar{v}\right).
\end{align}
%for $t\ge T_{ij}$. where, with some abuse of notation, the value of $\bar v\cdot(t-T_{ij})$ is imposed to take zero in the time domain $0\le t\le T_{ij}$.

Let $\Delta q_i=q_i-\bar{v}\cdot t$, and equation (\ref{eq8}) can be reformulated as
\begin{align}
\label{eq9}
\Delta\dot{q}_i=&-\Sigma_{j\in\mathcal{N}_i}w_{ij}\left[\Delta q_i-\Delta q_j(t-T_{ij})\right]\nn\\&+\left(1+\Sigma_{j\in\mc{N}_i}w_{ij}T_{ij}\right)\left({v}_i-\bar{v}\right)+s_i
\end{align}
where $\Delta q_j(t-T_{ij})=-\bar v\cdot(t-T_{ij})$ for $0\le t< T_{ij}$.

We propose the control law for the $i$-th mechanical system as
\be
\label{eq10}
\tau_i=Y_i(q_i,\dot{q}_i,\dot{q}_{r,i},\ddot{q}_{r,i})\hat{a}_i-K_i s_i
\ee
where $K_i$ is a symmetric positive definite matrix, and $\hat{a}_i$ is the estimate of the unknown parameter $a_i$, which is updated by the following adaptation law
\be
\label{eq11}
\dot{\hat{a}}_i=-\Gamma_i Y_i^T(q_i,\dot{q}_i,\dot{q}_{r,i},\ddot{q}_{r,i})s_i
\ee
where $\Gamma_i$ is a symmetric positive definite matrix.

\emph{Remark 2:} The adaptive controller (\ref{eq10}), (\ref{eq11}) is actually the well-known Slotine and Li adaptive control \cite{Slotine1987_IJRR} with new reference velocity and reference acceleration that take into account the interaction among the mechanical agents.

Substituting the control law (\ref{eq10}) into the dynamics (\ref{eq2}) yields
\be
\label{eq12}
M_i(q_i)\dot{s}_i+C_i(q_i,\dot{q}_i)s_i=-K_i s_i+Y_i(q_i,\dot{q}_i,\dot{q}_{r,i},\ddot{q}_{r,i})\Delta a_i
\ee
where $\Delta a_i=\hat{a}_i-a_i$ is the parameter estimation error.

The properties of the mechanical network can be adequately described by the following system
\be
\label{eq13}
\begin{cases}
\overbrace{\Delta\dot{q}_i=-\Sigma_{j\in\mathcal{N}_i}w_{ij}\left[\Delta q_i-\Delta q_j(t-T_{ij})\right]}^{\Psi}\\ \quad\qquad+\left(1+\Sigma_{j\in\mc{N}_i}w_{ij}T_{ij}\right)\left({v}_i-\bar{v}\right)+s_i,\\
\dot{{v}}_i=-\Sigma_{j\in \mathcal{N}_i}b_{ij}\left[{v}_i-{v}_j(t-T_{ij})\right],\\
M_i(q_i)\dot{s}_i+C_i(q_i,\dot{q}_i)s_i=-K_i s_i+Y_i(q_i,\dot{q}_i,\dot{q}_{r,i},\ddot{q}_{r,i})\Delta a_i, i\in\mathcal{V}.
\end{cases}
\ee

 The above system is cascaded in that the delay-dependent linear system $\Psi$ in the first subsystem is driven by the signals $v_i$ and $s_i$ generated by the lower two subsystems. The key challenge in derivation of the properties of the above system lies in the analysis of the first subsystem, which is mainly caused by the communication delays between the agents. To this end, applying the standard Laplace transformation to the first subsystem in (\ref{eq13}) gives
 \begin{align}
 \label{eq14}
 p\Delta Q_i(p)-\Delta q_i(0)=&-\Sigma_{j\in\mathcal{N}_i}w_{ij}\Big[\Delta Q_i(p)-e^{-T_{ij}p}\Delta Q_j(p)+\overbrace{\int_0^{T_{ij}}\bar v\cdot (t-T_{ij})e^{-pt}dt}^{\Phi_{ij}(p)}\Big]\nn\\
 &+\left(1+\Sigma_{j\in\mc{N}_i}w_{ij}T_{ij}\right)\Delta V_i(p)+S_i(p)
 \end{align}
 where $p$ denotes the Laplace variable, and $\Delta Q_i(p)$, $\Delta V_i(p)$ and $S_i(p)$ denote the Laplace transforms of $\Delta q_i$, $\Delta v_i=v_i-\bar{v}$ and $s_i$, respectively. The term $\Phi_{ij}(p)$ appears in (\ref{eq14}) since $\Delta q_j(t-T_{ij})=-\bar v\cdot(t-T_{ij})$ for $0\le t< T_{ij}$, and from the form of $\Phi_{ij}(p)$, we can regard it as the Laplace transform of a signal $\phi_{ij}(t)$ which is defined as $\phi_{ij}(t) =\bar{v}\cdot(t-T_{ij})$  for $0\le t<T_{ij}$ and $\phi_{ij}(t)=0$ for $t\ge T_{ij}$. For conciseness, define $\lambda_i(t)=\Sigma_{j\in\mathcal{N}_i}w_{ij}\phi_{ij}(t)$, $i=1,2,\dots,n$.

 Next, define $\lambda(t)=\left[\lambda_1^T(t),\lambda_2^T(t),\dots,\lambda_n^T(t)\right]^T$, and let $\Lambda(p)$ denote the Laplace transform of $\lambda(t)$. Furthermore, let $\Delta Q(p)=\left[\Delta Q_1^T(p),\Delta Q_2^T(p),\dots,\Delta Q_n^T(p)\right]^T$, $\Delta q(0)=\big[\Delta q_1^T(0),\Delta q_2^T(0),\dots,\Delta q_n^T(0)\big]^T$, $\Delta V(p)=\big[\Delta V_1^T(p),\Delta V_2^T(p),\dots,\Delta V_n^T(p)\big]^T$ and $S(p)=\left[S_1^T(p),S_2^T(p),\dots,S_n^T(p)\right]^T$. Then, using Kronecker product \cite{Brewer1978_TCS}, we can formulate equation (\ref{eq14}) at the velocity level, i.e.,
  \begin{align}
  \label{eq15}
 \Delta Q_v(p)=&\big[\overbrace{(pI_n+\mathcal{D}_w-\mathcal{W}_T(p))^{-1}}^{G(p)}\otimes I_m\big]\nn\\&\times\Big\{-\left[(\mathcal{D}_w-\mathcal{W}_T(p))\otimes I_m\right]\Delta q(0)\nn\\&+p\big[\underbrace{-\Lambda(p)+(\mathcal{D}_T\otimes I_m)\Delta {V}(p)+S(p)}_{\Omega(p)}\big]\Big\}
 \end{align}
where $\Delta Q_v(p)=p\Delta Q(p)-\Delta q(0)$ is the Laplace transform of $\Delta\dot{q}$, the delay-dependent matrix $\mathcal{W}_T(p)=\left[w_{ij}e^{-T_{ij}p}\right]$, the matrix $\mathcal{D}_w=\text{diag}\left[\Sigma_{j\in\mathcal{N}_i}w_{ij},i=1,2,\dots,n\right]$, and the matrix $\mathcal{D}_T=\text{diag}\left[1+\Sigma_{j\in\mathcal{N}_i}w_{ij}T_{ij},i=1,2,\dots,n\right]$. Note that $\Omega(p)$ is actually the Laplace transform of the signal $\omega(t)=-\lambda(t)+\left(\mathcal{D}_T\otimes I_m\right)\Delta v(t)+s(t)$.

We are presently ready to formulate the following theorem.

\emph{Theorem 1:} The adaptive controller (\ref{eq10}), (\ref{eq11}) with $v_i$ generated by the distributed observer (\ref{eq4}) ensures the second-order consensus of the networked mechanical systems on digraphs containing a spanning tree irrespective of the communication delays, i.e., $q_i(t)-q_j(t)\to 0$ and $\dot{q}_i(t)\to \bar{v}$ as $t\to\infty$, $\forall i,j\in\mathcal{V}$.

\emph{Proof:} Following \cite{Slotine1987_IJRR,Ortega1989_Aut}, we consider the Lyapunov-like function candidate for the third subsystem in (\ref{eq13}) $V_i=\frac{1}{2}s_i^T M_i(q_i)s_i+\frac{1}{2}\Delta a_i^T \Gamma_i^{-1}\Delta a_i$, and exploiting \emph{Property 2}, we obtain $$\dot{V}_i=-s_i^T K_i s_i\le 0$$
which gives the result that $s_i\in L_2\cap L_\infty$ and $\hat{a}_i\in L_\infty$, $\forall i\in\mathcal{V}$.

Let us rewrite equation (\ref{eq15}) as
  \begin{align}
  \label{eq16}
 \Delta Q_v(p)=\left[G(p)\otimes I_m\right]\Big\{\overbrace{-\left[(\mathcal{D}_w-\mathcal{W}_T(p))\otimes I_m\right]\Delta q(0)+\omega(0)}^\text{initial-condition-dependent term}+p\Omega(p)-\omega(0)\Big\}.
 \end{align}
From \cite{Nuno2011_TAC}, we know that all the poles of $G(p)$ excluding the simple zero pole are in the open left half plane (LHP), and therefore, $G(p)$ is iBIBO stable \cite{Wang2013b_TAC}. Next, we show the boundedness of $\Delta\dot{q}$ using a procedure similar to \cite{Wang2013b_TAC}. From the standard linear system theory, the initial-condition-dependent term in (\ref{eq16}), passing through the marginally stable system $G(p)$, results in a bounded output. Note that $p\Lambda(p)-\lambda(0)$, $p\Delta V(p)-\Delta v(0)$ and $pS(p)-s(0)$ represent the Laplace transforms of $\dot\lambda(t)$, $\Delta\dot{v}(t)$ and $\dot{s}(t)$, respectively, and the integrations of the three quantities are $\int_0^t \dot \lambda(r)dr=\lambda(t)-\lambda(0)\in L_\infty$, $\int_0^t\Delta\dot{v}(r)dr=\Delta v(t)-\Delta v(0) \in L_\infty$ and $\int_0^t \dot{s}(r)dr=s(t)-s(0)\in L_\infty$. Therefore, the input $p\Omega(p)-\omega(0)$ [i.e., the Laplace transform of $\dot \omega(t)=-\dot \lambda(t)+\left(\mathcal{D}_T\otimes I_m\right)\Delta\dot v(t)+\dot s(t)$] is integral-bounded, which must give a bounded output after passing through $G(p)$ since $G(p)$ is iBIBO stable. From the superposition principle of linear systems, we obtain the boundedness of $\Delta Q_v(p)$, i.e., $\Delta \dot q\in L_\infty$.

From the first subsystem in (\ref{eq13}), we have $\Sigma_{j\in\mathcal{N}_i}w_{ij}\left[\Delta q_i-\Delta q_j(t-T_{ij})\right]\in L_\infty$ since $\Delta v_i$ and $s_i$ are both bounded, $\forall i\in\mathcal{V}$.
We also obtain $\Sigma_{j\in\mathcal{N}_i}w_{ij}\left[q_i-q_j(t-T_{ij})\right]=\Sigma_{j\in\mathcal{N}_i}w_{ij}\left[\Delta q_i-\Delta q_j(t-T_{ij})\right]+\Sigma_{j\in\mathcal{N}_i}w_{ij}T_{ij}\bar v\in L_\infty $, and then, from (\ref{eq6}), we obtain the boundedness of $\dot{q}_{r,i}$, $\forall i\in\mathcal{V}$. Therefore, $\dot{q}_i=s_i+\dot{q}_{r,i}\in L_\infty$, $\forall i\in\mathcal{V}$. The boundedness of $\dot{v}_i$ can be straightforwardly derived from equation (\ref{eq4}), and hence, $\ddot{q}_{r,i}$ is bounded, $\forall i\in\mathcal{V}$. From (\ref{eq12}), we get the boundedness of $\dot{s}_i$ since $M_i(q_i)$ is uniformly positive definite (by \emph{Property 1}), giving rise to the boundedness of $\ddot{q}_i$, $\forall i\in\mathcal{V}$. Therefore, $s_i$ is uniformly continuous, and from \cite{Lozano2000_Book} (p. 117), we obtain $s_i\to 0$ as $t\to\infty$, $\forall i\in\mathcal{V}$. From the \emph{final value theorem}, we have $\lim_{p\to0}pS(p)=\lim_{t\to\infty}s(t)=0$, and we also have $\lim_{p\to0}p\Delta V(p)=\lim_{t\to\infty}\Delta v(t)=0$ (by \emph{Lemma 2}) and $\lim_{p\to 0}p \Lambda(p)=\lim_{t\to\infty}\lambda(t)=0$. Therefore, $\lim_{p\to 0}p\Omega(p)=0$. From \cite{Wang2013b_TAC}, we know that
\be
\lim_{p\to 0}pG(p)=\sigma_S  1_n\gamma^T
\ee
where $\sigma_S=\frac{1}{1+\Sigma_{i=1}^n\Sigma_{j\in\mathcal{N}_i}\gamma_i w_{ij} T_{ij}}$. Therefore, from (\ref{eq16}), invoking the \emph{final value theorem} and using 2) of \emph{Lemma 1}, we get
\be
\label{eq18}
\lim_{t\to \infty}\Delta \dot{q}(t)=\sigma_S \left[\left(1_n \gamma^T \mathcal{L}_w\right)
\otimes I_m\right]\Delta q(0)=0
\ee
which directly gives the result that $\dot q_i(t)\to\bar v$ as $t\to\infty$, $\forall i\in\mathcal{V}$.
Using (\ref{eq18}), we obtain $\Delta q_j(t)-\Delta q_j(t-T_{ij})=\int_0^{T_{ij}}\Delta\dot{q}_j(t-r)dr\to 0$ as $t\to\infty$, $\forall j\in \mathcal{N}_i,i\in\mathcal{V}$. From the first subsystem in (\ref{eq13}), we have
\be
0=-\left(\mc{L}_w\otimes I_m\right)\Delta q(\infty).
\ee
Note that $\Delta q(t)=q(t)-\left(1_n\otimes I_m\right)\bar{v}\cdot t$, and since $\mathcal{L}_w 1_n=0$ (by \emph{Lemma 1}), we obtain
\begin{align}
\lim_{t\to\infty}-\left(\mc{L}_w\otimes I_m\right)\Delta q(t)=&\lim_{t\to\infty}-\left(\mc{L}_w\otimes I_m\right)q(t)+\left(\mc{L}_w\otimes I_m\right)\left[\left(1_n\otimes I_m\right)\bar{v}t\right]\nn\\
=&{-\left(\mc{L}_w\otimes I_m\right)q(\infty)}=0.
\end{align}
From \emph{Lemma 1}, we know that $\mathcal{L}_w$ has a unique basis vector $1_n$, and thus we have $q_i(t)-q_j(t)\to 0$ as $t\to\infty$, $\forall i,j\in\mc{V}$. \hfill {\small $\blacksquare$}

\emph{Remark 3:} In the special/reduced case that the agents are governed by the double-integrator dynamics $\ddot{q}_i=\tau_i$, the proposed second-order consensus algorithm gives
\be
\begin{cases}
\dot{{v}}_i=-\Sigma_{j\in \mathcal{N}_i}b_{ij}\left[{v}_i-{v}_j(t-T_{ij})\right], v_i(0)=\dot q_i(0),\\
\ddot q_i=-\Sigma_{j\in\mathcal{N}_i}w_{ij}\left[\dot q_i-\dot q_j(t-T_{ij})\right]-k\Sigma_{j\in\mathcal{N}_i}w_{ij}\left[ q_i- q_j(t-T_{ij})\right]\\
+\underbrace{\left(1+\Sigma_{j\in\mathcal{N}_i}w_{ij}T_{ij}\right)\left(\dot v_i+k v_i\right)-k \dot q_i}_{\Pi}, \forall i\in{\cal V},
\end{cases}
\ee
where $k>0$ is a positive scalar. From \emph{Theorem 1}, we immediately obtain that the second-order consensus is realized, i.e., $\dot q_i(t)\to \frac{1}{1+\Sigma_{k=1}^n\Sigma_{j\in{\cal N}_k}\gamma_k b_{kj}T_{kj}}\Sigma_{k=1}^n\gamma_k\dot{q}_k(0) $ and $q_i(t)-q_j(t)\to0$ as $t\to\infty$, $\forall i,j\in {\cal V}$. The major difference between the protocol here and those in the literature (see, e.g., \cite{Wang2006_TAC}) is the additional delay-compensation term $\Pi$.

\section{Leader-Follower Second-Order Consensus}

 It is known that the leader-follower consensus can be considered as a special case of the consensus without a leader (see, e.g.,  \cite{Ren2008_Aut,Ren2008_Book}). Specifically, the application of this idea/viewpoint here results in the convenient extension from our previous consensus scheme to the one for achieving delay-robust leader-follower consensus for mechanical systems with a virtual constant-velocity leader (denoted by vertex 0), as will be demonstrated below. Let $q_L$ and $\dot q_L$ denote the position and velocity of the leader, and suppose that the leader's velocity $\dot{q}_L$ is constant and that the graph  among the virtual leader and the $n$ mechanical followers (denoted by $\mathcal{G}^\ast$) contains a spanning tree rooted at vertex 0.

In the leader-follower case, the distributed velocity observer (\ref{eq4}) becomes
\begin{align}
\label{eq21}
\dot{{v}}_i=-\Sigma_{j\in \mathcal{N}_i}b_{ij}\left[{v}_i-{v}_j(t-T_{ij})\right]-b_{i0}\left[ v_i-\dot{q}_L(t-T_{i0})\right]
\end{align}
where the weight $b_{i0}$ is defined as $b_{i0}>0$ if the $i$-th mechanical agent can directly access the information of the leader and $b_{i0}=0$ otherwise, $\forall i\in\mathcal{V}$. In the case $t\ge T_{i0}$, since $\dot q_L(t-T_{i0})=\dot q_L$, we can rewrite equation (\ref{eq21}) as
\be
\label{eq22}
\dot{{v}}_i=-\Sigma_{j\in \mathcal{N}_i}b_{ij}\left[{v}_i-{v}_j(t-T_{ij})\right]-b_{i0}\left[ v_i-\dot{q}_L\right], \forall i\in\mathcal{V}.
\ee
It is obvious that the distributed observer (\ref{eq22}) has the same convergent properties as (\ref{eq21}) except in the initial stage.

As to the control strategy for the $i$-th mechanical system, we can still employ the adaptive control action (\ref{eq10}), (\ref{eq11}), yet, the reference velocity $\dot{q}_{r,i}$ in (\ref{eq6}) needs to be redefined as
\begin{align}
\label{eq23}
\dot{q}_{r,i}=&\left(1+w_{i0}T_{i0}+{\Sigma_{j\in{\cal N}_i}w_{ij}T_{ij}}\right){v}_i-\Sigma_{j\in{\cal N}_i} w_{ij}\left[q_i-q_j(t-T_{ij})\right]\nn\\&-w_{i0}\left[q_i-q_L(t-T_{i0})\right]
\end{align}
where the weight $w_{i0}$ is defined as $w_{i0}>0$ if the $i$-th mechanical agent can directly access the information of the leader and $w_{i0}=0$ otherwise, and $T_{i0}$ is the constant communication delay from the leader to agent $i$.

\emph{Corollary 1:} The distributed observer (22) ensures that $v_i(t)\to \dot{q}_L$ as $t\to\infty$, $\forall i\in\mathcal{V}$ provided that the graph among the virtual leader and the $n$ mechanical followers contains a spanning tree.

\emph{Proof:} To demonstrate the convergent property of (\ref{eq22}), we extend the distributed observer (\ref{eq22}) to include the dynamics of the virtual leader, which gives
\be
\label{eq24}
\begin{cases}
\ddot{q}_L=0,\\
\dot{{v}}_i=-\Sigma_{j\in \mathcal{N}_i}b_{ij}\left[{v}_i-{v}_j(t-T_{ij})\right]-b_{i0}\left[ v_i-\dot{q}_L\right], i\in\mathcal{V}.
\end{cases}
\ee
If we take the followers $1,2,\dots,n$ and the virtual leader $0$ as a whole (which, in fact, forms a graph containing a spanning tree with vertex 0 as the unique root), then, the leader-follower case becomes a special one of the distributed observer (\ref{eq4}). Then, from \emph{Lemma 2} and based on 3) in \emph{Lemma 1}, it is straightforward to prove \emph{Corollary 1}, and in fact $\bar v$, in this case, is equal to $\dot{q}_L$. \hfill{\small $\blacksquare$}

\emph{Corollary 2:} The adaptive controller (\ref{eq10}), (\ref{eq11}) with the reference velocity $\dot q_{r,i}$ defined by (\ref{eq23}) ensures the leader-follower consensus if the graph among the $n$ mechanical followers and the virtual leader has a spanning tree, i.e., $q_i(t)-q_L(t)\to 0$ and $\dot{q}_i(t)-\dot{q}_L\to 0$ as $t\to\infty$, $\forall i\in\mathcal{V}$.

\emph{Proof:} Similar to the proof of \emph{Corollary 1}, let us take the followers $1,2,\dots,n$ and the virtual leader $0$ as a whole, then, the leader-follower consensus problem can be considered as a special case of the previous second-order consensus problem. Thus, the result in \emph{Corollary 2} follows. \hfill{\small $\blacksquare$}

\emph{Remark 4:} \emph{Corollary 1} can be equivalently stated as that the system (\ref{eq22})
or (let $\Delta v_i=v_i-\dot{q}_L$)
\be
\label{eq25}
\Delta \dot{v}_i=-\Sigma_{j\in \mathcal{N}_i}b_{ij}\left[\Delta {v}_i-\Delta {v}_j(t-T_{ij})\right]-b_{i0}\Delta v_i, i=1,2,\dots,n,
\ee
is asymptotically stable provided that the graph among the $n$ followers and the virtual leader contains a spanning tree. One related result appears in \cite{Nuno2011_TAC}, which ensures the asymptotic stability of a system like (\ref{eq25}) under a relatively strong assumption that all followers are able to directly access the information of the leader (i.e., $b_{i0}>0$, $\forall i\in\mathcal{V}$). Another related result is given in \cite{Wang2006_TAC}, which demonstrates the delay-independent stability of (\ref{eq22}) under the condition that the graph topology is a ring (with identical weight) or undirected.  If the communication delays $T_{ij}$ are uniform, the result in \emph{Corollary 1} would coincide with that in \cite{Liu2012_CCC,Liu2012_AAS,Meng2011_SMC} except that the result here does not employ the normalized weight as in \cite{Liu2012_CCC,Liu2012_AAS,Meng2011_SMC}.

\emph{Remark 5:} The delay-robust leader-follower scheme here, as a special case of our main result, allows the communication delays to be arbitrary finite constants, which is contrary to \cite{Peng2009_PhysicaA,Zhu2010_Aut,Yang2010_EJC,Meng2010_EJC,Meng2011_SMC}, and additionally distributed, in contrast with \cite{Zhu2010_Aut}. In addition, the agents considered here are governed by nonidentical nonlinear mechanical dynamics with parametric uncertainties while the models of the agents considered in \cite{Peng2009_PhysicaA,Zhu2010_Aut,Yang2010_EJC,Meng2010_EJC,Meng2011_SMC} are identical double-integrators. The results in \cite{Liu2012_AAS,Liu2012_CCC} consider the mechanical systems, yet, they cannot ensure asymptotic convergence under a dynamic leader, and moreover the communication delays are required to be uniform. The results in \cite{Mei2011_TAC,Mei2012_Aut} realize leader-follower second-order consensus of mechanical systems, yet, the communication delay is assumed to be absent.

\section{Simulation Results}

Let us illustrate the performance of the proposed second-order consensus scheme via simulations on six standard two-DOF planar robots. The interaction topology is shown in Fig 1, where the
cases with or without a leader are both plotted.
The gravitational torques of the six robots are not considered for simplicity, i.e., set $g_i(q_i)\equiv0$, $i=1,2,\dots,6$, and the physical parameters of the six robots are not listed for saving space. The sampling
period is set to be 5 ms.

\begin{figure}
\centering
%%----start of first figure----
\begin{minipage}[t]{1.0\linewidth}
\centering
\includegraphics[width=2.3in]{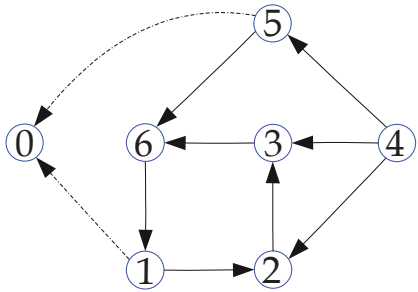}
\caption{The interaction graph among the mechanical agents with or without a leader (i.e., vertex 0)}\label{fig:side:a}
\end{minipage}%
\end{figure}

\subsection{Second-Order Consensus Without a Leader}

We first consider the case without a leader, i.e., vertex $0$ and the associated edges $(1,0)$ and $(5,0)$ are absent in Fig. 1. The communication delays among the agents, for simplicity, are set to be $T_{ij}=0.5\text{ s}$, $\forall i\in\mathcal{V},j\in\mathcal{N}_i$. The adjacency weights for the distributed velocity observer are determined as $b_{ij}=1.5$, $\forall i\in\mathcal{V},j\in\mathcal{N}_i$. The entries of the weighted adjacency matrix $\mathcal{W}$ are determined as $w_{ij}=1.0$, $\forall i\in\mathcal{V},j\in\mathcal{N}_i$. The controller parameters $K_i$ and $\Gamma_i$ are chosen as $K_i=40.0 I$ and $\Gamma_i=2.0I$, respectively, $i=1,2,\dots,6$. The initial parameter estimates are chosen as $\hat{a}_i(0)=0$, $i=1,2,\dots,6$. Simulation results are shown in Fig. 2 and Fig. 3 (to save space, only the first coordinate of the position/velocity of the robotic agent is plotted). From Fig. 2 and Fig. 3,  we see that the positions of the mechanical agents synchronize and their velocities converge to the scaled weighted average value $\bar{v}=\left[0.4286,-0.2857\right]^T$.

\subsection{Second-Order Consensus With a Leader}

Let us now consider the case of existence of a leader, and the communication interaction among the mechanical agents and the virtual leader is fully characterized by the topology given in Fig. 1, where vertex 0 represents the virtual leader.  The velocity of the virtual leader is set as $\dot {q}_L = \left[
 {1.5}, {2.0} \right]^T$, and its initial position is
set as $q_L (0) = \left[
 {0.5}, {0.5}\right]^T$. The weights
describing the relation between the followers and the leader are set as
$w_{10} = w_{50} = 1.0$, $w_{20} = w_{30} = w_{40} = w_{60} = 0.0$, $b_{10}
= b_{50} = 1.5$, $b_{20} = b_{30} = b_{40} = b_{60} = 0.0$. The communication delays
$T_{i0},i=1,5$ are set to be 0.5 s. The adjacency weights that describe the relation among the followers are set to be the same as the case without a leader, so are the communication delays among the followers. The controller parameters $K_i$ and $\Gamma_i$, and the initial parameter estimate $\hat{a}_i(0)$ are also chosen to be the same as the case without a leader, $i=1,2,\dots,6$.
Simulation results are shown in Fig. 4 and Fig. 5, from which, we see that
the proposed controller achieves the leader-follower asymptotic consensus, irrespective
of the communication delay.

\begin{figure}
\centering
%%----start of first figure----
\begin{minipage}[t]{0.4\linewidth}
\centering
\includegraphics[width=2.7in]{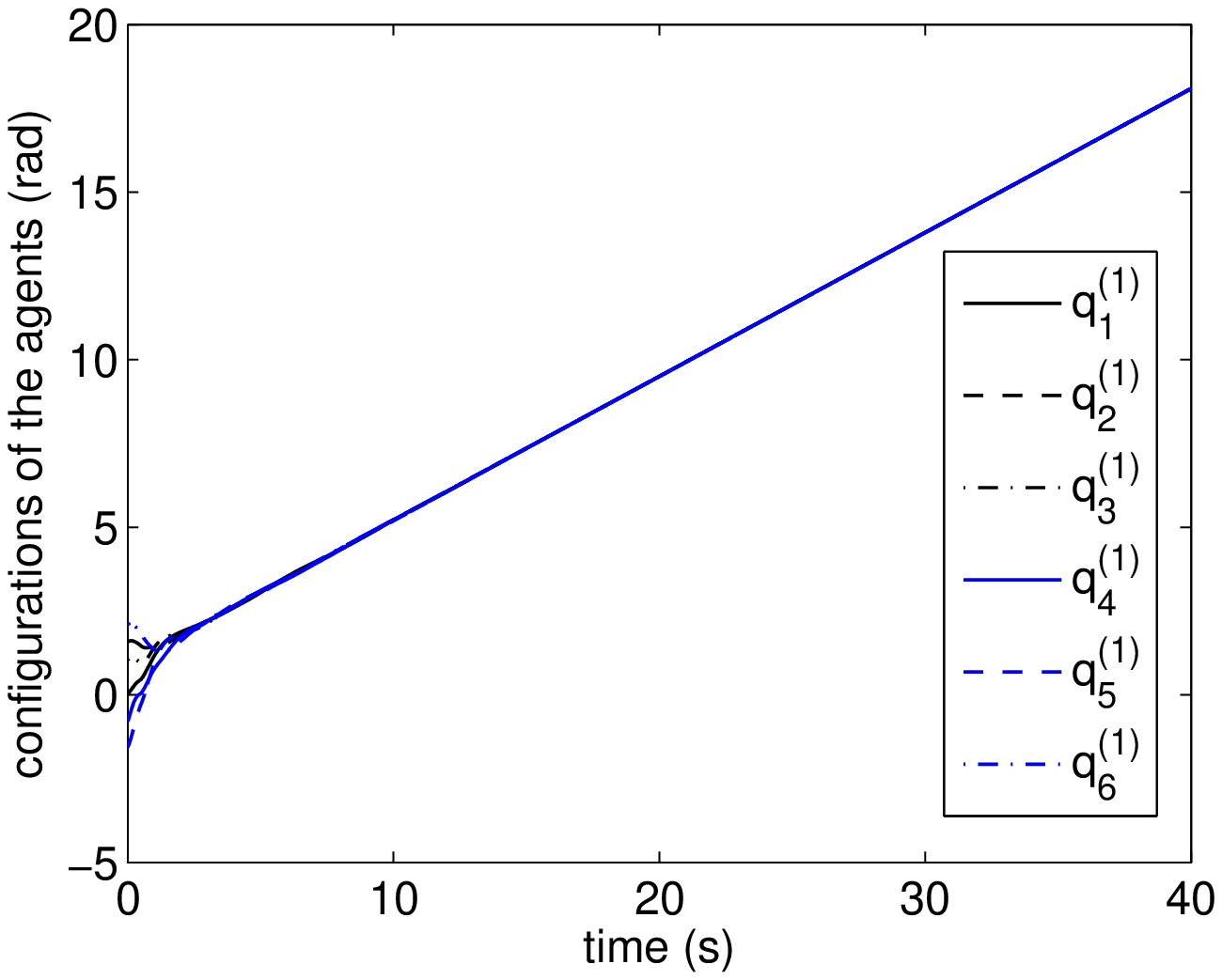}
\caption{Configuration variables of the robotic agents (the first coordinate)}\label{fig:side:a}
\end{minipage}%
\hspace{2cm}%
%%----start of second figure----
\begin{minipage}[t]{0.4\linewidth}
\centering
\includegraphics[width=2.7in]{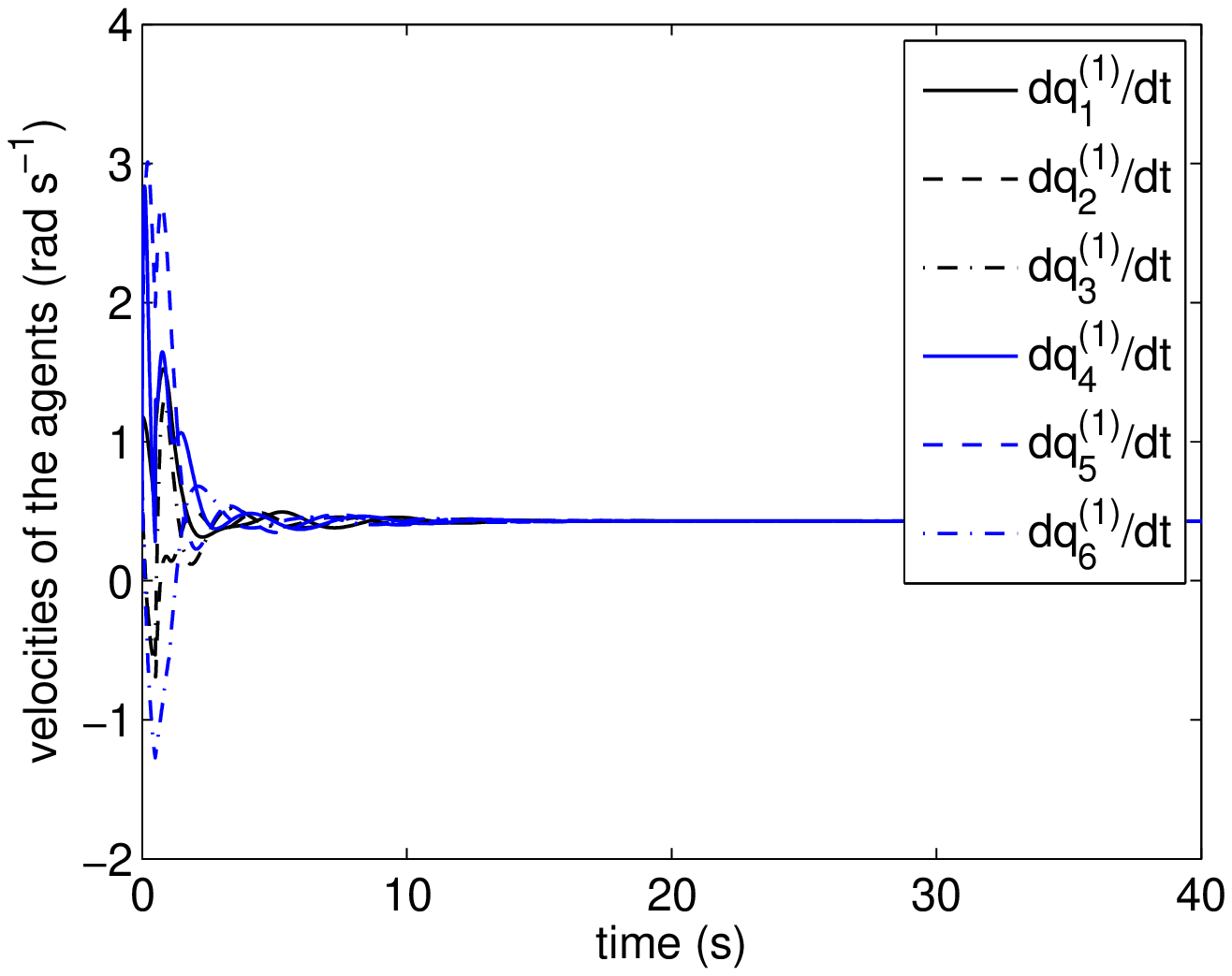}
\caption{Velocities of the robotic agents (the first coordinate)} \label{fig:side:b}
\end{minipage}
\end{figure}
\begin{figure}
\centering
%%----start of first figure----
\begin{minipage}[t]{0.4\linewidth}
\centering
\includegraphics[width=2.7in]{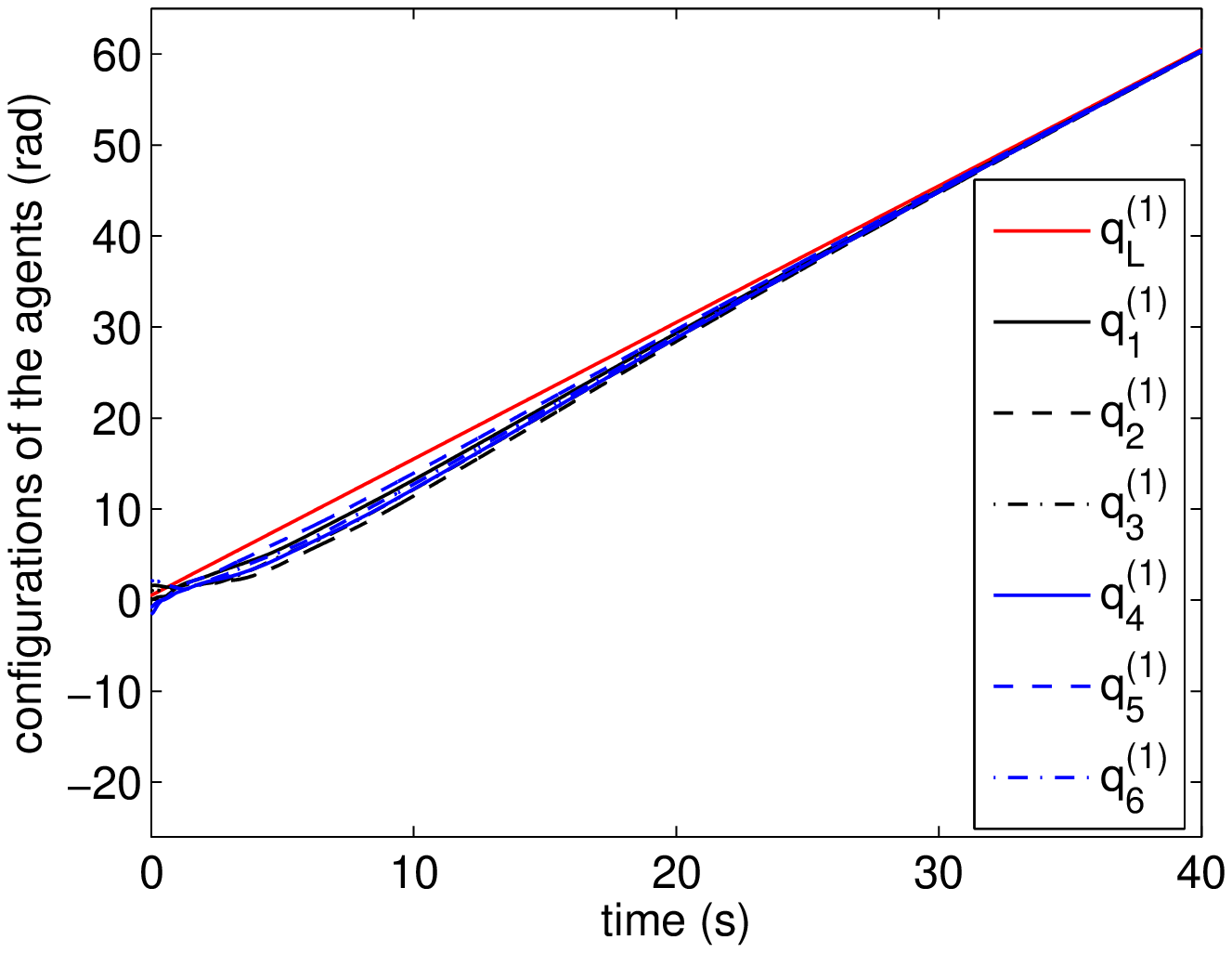}
\caption{Configuration variables of the robotic agents (the first coordinate)}\label{fig:side:a}
\end{minipage}%
\hspace{2cm}%
%%----start of second figure----
\begin{minipage}[t]{0.4\linewidth}
\centering
\includegraphics[width=2.7in]{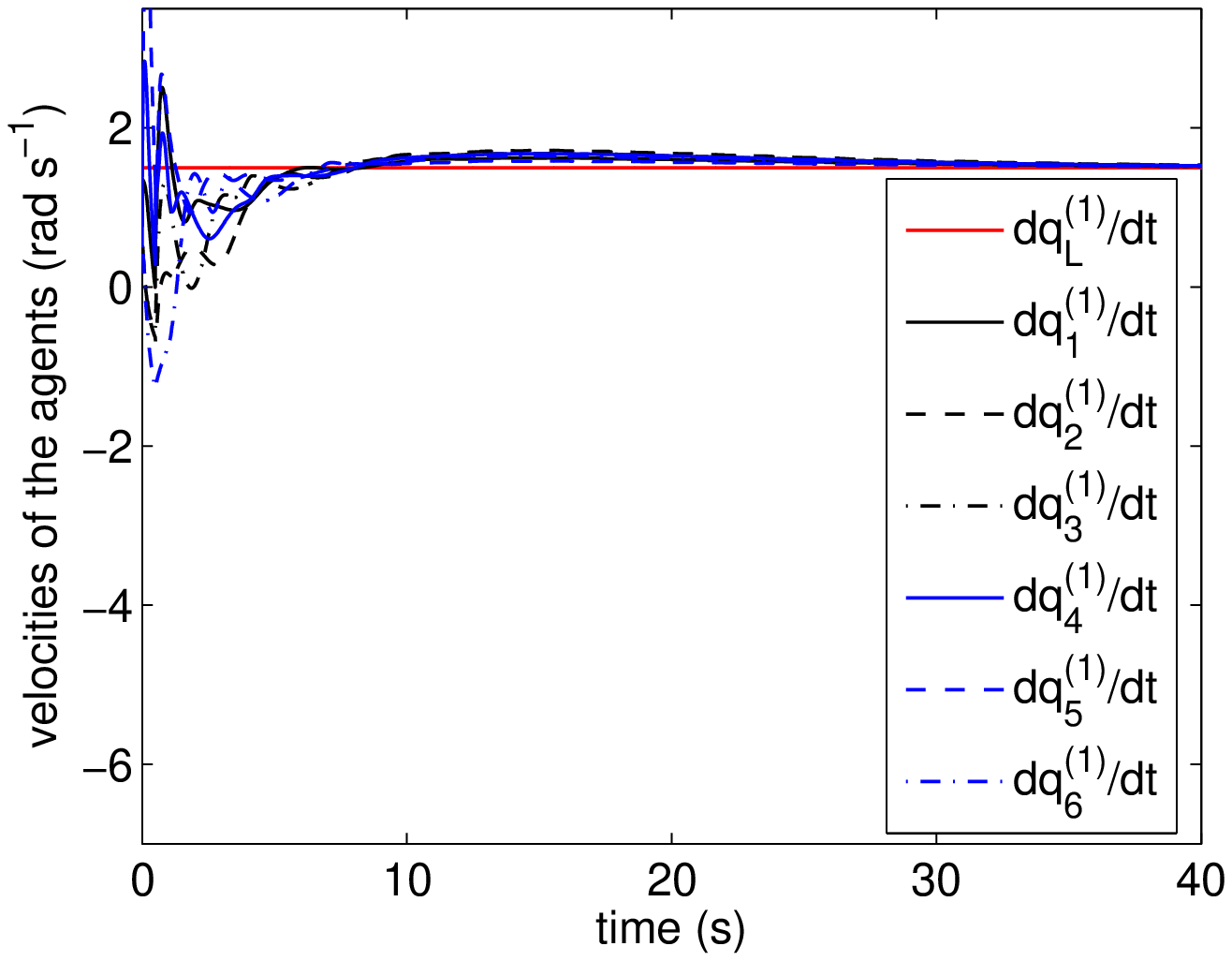}
\caption{Velocities of the robotic agents (the first coordinate)} \label{fig:side:b}
\end{minipage}
\end{figure}

\section{Conclusion}

In this paper, we have examined the second-order consensus problem for multiple mechanical systems on a directed graph and under nonuniform communication delays. An adaptive scheme, which are composed by an adaptive controller and a delay-robust distributed velocity observer, is proposed to achieve the goal of second-order consensus. Using the iBIBO stability analysis and the {final value theorem}, we show that the position/velocity synchronization errors between the mechanical agents asymptotically converge to zero, and in addition, the velocities of the mechanical agents converge to the scaled weighted average value of their initial ones. Then, we illustrate that the control scheme for the second-order leader-follower consensus problem with a constant-velocity leader is contained in the proposed delay-robust consensus framework. Simulation results are presented to demonstrate the performance of the proposed controllers.

% or


%\appendices

%\section{}

%\section*{Acknowledgment}


%\ifCLASSOPTIONcaptionsoff
%  \newpage
%\fi



\bibliographystyle{IEEEtran}
\bibliography{..//Reference_list_Wang}
%\begin{thebibliography}{1}
%
%
%\end{thebibliography}

%\begin{biography}[{\includegraphics[width=1in,height=1.25in,clip,keepaspectratio]{mshell}}]{Michael Shell}

%\end{IEEEbiography}


%\newpage



%\vfill


\end{document}